# Intrusion Prediction with System-call Sequence-to-Sequence Model

ShaoHua Lv, Jian Wang, YinQi Yang, JiQiang Liu
School of Computer and Information Technology
Beijing Jiaotong University, China
{16120401, wangjian, 17120486, jqliu}@bjtu.edu.cn

**Abstract.** The advanced development of the Internet facilitates efficient information exchange while also been exploited by adversaries. Intrusion detection system (IDS) as an important defense component of network security has always been widely studied in security research. However, research on intrusion prediction, which is more critical for network security, is received less attention. We argue that the advanced anticipation and timely impede of invasion is more vital than simple alarms in security defenses. General research methods regarding prediction are analyzing short term of system-calls to predict forthcoming abnormal behaviors. In this paper we take advantages of the remarkable performance of recurrent neural networks (RNNs) in dealing with long sequential problem, introducing the sequence-to-sequence model into our intrusion prediction work. By semantic modeling system-calls we build a robust system-call sequence-to-sequence prediction model. With taking the system-call traces invoked during the program running as known prerequisite, our model predicts sequence of system-calls that is most likely to be executed in a near future period of time that enabled the ability of monitoring system status and prophesying the intrusion behaviors. Our experiments show that the predict method proposed in this paper achieved well prediction performance on ADFA-LD intrusion detection test data set. Moreover, the predicted sequence, combined with the known invoked traces of system, significantly improves the performance of intrusion detection verified on various classifiers.

## 1 Introduction

Intrusion detection systems (IDS) [1] are critical components with respect to security defenses of the networks. Traditional IDS, whether signature-based or anomaly-based, adopts intrusion detection technology based on certain model patterns. Signature-based methods match the system behaviors against the known attack patterns, while anomaly-based techniques build the system behavior models through prior knowledge to identify deviations. Both the two methods possess vulnerabilities, signature-based IDS lack the ability to detect zero-day attack, while anomaly-based IDS receive criticisms for its inherent property of yielding high false alarm rate.

Analyzing system-call traces which invoked by the running programs to discriminate normal and abnormal behaviors was first introduced by Forrest *et al.* [2]. Former anomaly intrusion detection approaches which focus on characterizing normal activities were less stable for the erratic user behaviors. Then in their subsequent works they define normal profile using short sequences of system-calls. With neural networks gained state-of-the-art performance in dealing with computer vision and natural language processing problems, researchers recently introduce neural networks in system call modeling [3], while most of which are dealing the frequency features of individual system-calls. In other words, they did not consider the semantic meaning of system-calls. More importantly, those methods unable have the ability to identify the intrusion intention of the malicious attacks. The concern of system-calls has been a vital research target of intrusion detection and the system-call sequences imply the program characteristics. Our work will implement on the system-call sequences.

In real network environment, the normal operation of the host nodes is vital to the system, any hostile attacks that destruct the service running will result in huge amount of asset loss. Therefore, it is not enough to alert the intrusion activities only. It needs method to monitor the program running status and predict the possible intrusion behaviors which would thread the stability and health of the networks.

To solve the above problem, in this paper we construct an end-to-end neural network [4] system-call prediction [5] model, which predicts the follow-up invoked system-calls based on the system-call traces that have been executed during the malicious process running. The predicted sequence is that the attack program is most likely to execute within a certain period of time.

System-call prediction is hard work for the arbitrary execute of those running programs, previous works use dynamic Bayesian network or hidden Markov model to predict sliding windowed system-calls. Those algorithms have long been criticized for their difficulty in calculating and the poor performance which caused by their short dependency on previous system-calls.

Much like human communicate with each other through languages constructed with certain grammar and syntax, processes achieve intention with specific system-call sequences. With the high similarity between natural human languages and system-call sequences, we employ the language model [6] to model the system calls. Because RNN is suitable for solving sequential problems, and it has significant performance in natural language processing research, such as Q&A and machine translation, we take advantages of the RNNs' sequence-to-sequence framework [7] to construct our system-call sequence prediction model.

The advantage of the method proposed in this paper lies in that it not only detects the malicious system-call sequence in real time, also predicts the possible system-call sequence at the next moment based on the given context. Most important of all, our work models the semantic meaning of system calls so that we can analyze sequences in sentence level. Then we can extract the abstract features of the system-call traces, thereby enhancing the accuracy and robustness of the anomaly detection. The contributions of this paper are as follows:

1) Employing RNNs language model to model system-calls which addresses the long temporal dependency problem, then applying sequence-to-sequence framework to generate predicted sequences conditioned on the given system-call context.
2) Improving the performance of anomaly intrusion detection system with lower false alarm rate by using the predicted sequences.
3) Addressing the defect of unable to make intrusion prediction and identifying the intrusion intention of the malicious adversary by constructing the system-call sequence predict model.

The follow-up parts of this article will be organized as follows: Section II. Describes related work, Section III. Explains our prediction model in detail, Section IV. Experiments and analysis of results and Section V. Conclusions.

## 2   Related Work

### 2.1   Intrusion Detection and Prediction

Forrest *et al.* [1, 2] first introduced system calls into anomaly intrusion system research works. In their proposed methods, 'normal' was defined by short-range correlations in a process's system call. Later they detected intrusions with short system-call sequences. Neural networks have recently been applied to improve the performance of intrusion detection systems [8, 9, 10]. Staudemeyer *et al.* [9] applied LSTM neural network to intrusion detection and achieved preferable results on KDD cup99 dataset. Kim *et al.* [10] further introduced the RNNs language model into anomaly intrusion detection system construct and outperformed previous methods. However, as fore mentioned, those intrusion detection technologies can only determine the intrusion activities that have occurred, nevertheless have no ability to neither forewarn the forthcoming invasion nor predict the subsequence of the malwares.

With respect to intrusion prediction, Feng *et al.* [11] proposed dynamic Bayesian networks to predict the subsequent actions and analyze intentions of the intruders based on system-calls. Zhang *et al.* [12] developed a hidden semi-Markov model method for predicting the anomaly events and the intentions of possible intruders to a computer system based on the observation of system-call sequences. Li *et al.* [14] constructed a network attack graph based on correlation analysis of large intrusion alerts then calculated the probability of each attack scenario in the graph node, that is, the probability of each attack scenario as a precursor to a future attack scenario. Most researchers use Hidden Markov Models [12, 13] to associate observed status with aggressive behaviors and use the transition probabilities to predict the next stage of intrusion behavior. There are quite a number of weaknesses in this approach. The first is the limited dependence on the sequence. The information on which the next attack is based are only with short steps. This will inevitably lose certain important information that contribute to the prediction result. Second, HMM can easily converge to local optimal point, which makes it difficult to find the accuracy prediction results.

## 2.2 RNNs Language Model

Recurrent neural networks have natural advantages in dealing with sequential data for its introducing cycles in their computational graph. The recurrent neural network contains a hidden state whose value is determined by the current time step input and the hidden state value of the previous time.

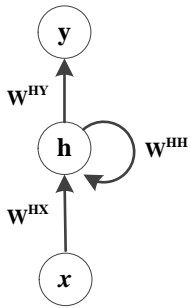

**Fig. 1.** the recurrent neural network model

Therefore, the output of each time step of the RNNs contains the information about the past inputs. If the recurrent neural network is expanded according to time series, it can be seen as a neural network whose depth is the length of the sequential inputs, and inputs of each hidden layer are the corresponding time step of original sequential data and the output of the previous layer. Fig. 1 shows the simple structure of recurrent neural networks.

Vanilla recurrent neural networks are trained with Backpropagation through time (BPTT) [15] algorithm, and it can in principle memorize information for infinitely long sequence. However, training models with long inputs will suffer from the problems of gradient vanishing or gradient explosion [16, 17]. The long short-term memory (LSTM) [18] and gated recurrent unit (GRU) [19] memory networks have the ability to remember important information of the long sequence by training the parameters of the input gate, forget gate, and output gate, which solving the problem of long-term dependency to a certain extent. Compared with LSTM, GRU has a simpler structure and can reduce the parameters needed to train. Chung *et al.* [20] evaluated GRU in various data set to found that it achieved the same performance against LSTM. For this reason, in this paper we choose GRU as the RNNs implement unit in our system-call prediction model.

The language model [21] can be expressed as the probability of a word sequence be a sensible sentence, and the language model can be represented by the conditioned probability of the next word given all the previous ones. The formula is as follows:

$$\hat{p}(w_1^T) = \prod_{t=1}^{T} \hat{p}(w_t | w_t^{t-1}) \tag{1}$$

where $w_t$ stands for the $t_{th}$ word in the sequence, and $T = (w_1, w_2, ..., w_t)$ is the given sentence. The recurrent neural network language model [22] has achieved great success in the fields of text generation and machine translation. It is an important part of the sequence-to-sequence model and determines the semantic rationality of the predicted word sequence. Taking the advantage of the generality nature of RNNs language model, we employ it to the system-call generation situation in this paper.

## 3 System-call Sequence-to-Sequence Model

### 3.1 Overview of Our Approach

The System-call Sequence-to-sequence model is modeling system-calls with sequence-to-sequence approach. Fig. 2 shows the overview of our proposed mothed, there are mainly two parts of our method. First, to fulfill our object of solving the shortcomings mentioned above of anomaly detection systems as well as handling the short dependency problems in existed system-call prediction algorithms, we proposed a sequence-to-sequence recurrent system-call prediction model. Second, to improve the performance of baseline intrusion detection classifiers, we combined the predicted sequence as a supplementary part with invoked sequence to form an extended input. As show in Fig. 2, the predicted sequence supplies the classifiers extra information to enhance the ability to define the system-call sequence nature.

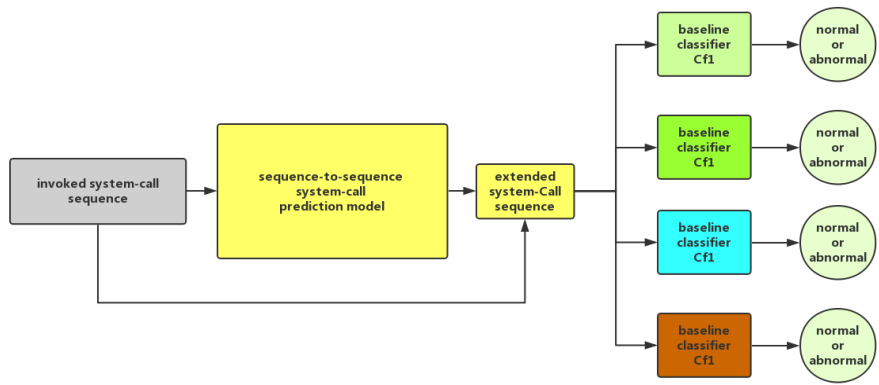

**Fig. 2.** Overview of the proposed method.

### 3.2 Prediction Model

We treat the system-call traces generated during program execution as language sentences which used by the process to communicate with operation system, we apply the RNNs language model as the system-call generator to ensure that we produce semantically reasonable sequences. By introducing the sequence-to-sequence framework which well performed in natural language processing, we feed the input system-call sequence as the context information to generate the correct prediction output related to the input sequence.

The sequence-to-sequence model is essentially a RNNs implementation of the encoder-decoder framework [23], and it can be viewed as a sentence-to-sentence mapping. Machine translation is a mapping of one language sentence to another, and the Q&A system is sentence mapping between the same languages.

The idea of encoder and decoder is a simulation of the way that human thinking. When answering a question, we usually understand the question first, and then construct the answer statement based on the content we understanding. In the same way, we divide a system-call trace into two segments according to its invoking order. We can treat the first part, which we call it *source* sequence, as a question statement in a Q&A system, and the second part, called *target*, as an answer statement. In fact, different proportions of the *source* and *target* can be set to divide the traces. We adopt dynamic proportional segmentation, that is, the ratio of input sequence length to output sequence length changes dynamically. Based on this idea, we construct the training data set of the sequence-to-sequence model. We divide 80% of the data set for training and 20% for evaluating.

Generally, a vocabulary of words is needed to accommodate all the words that may appear in the sentence. Similarly, we need to construct corresponding system-call word vocabulary when using the sequence-to-sequence prediction model. It is certain that the total number of system calls in the host is determined, assuming that is *m*. We define *S* as a set of all system calls provided in an operating system. We number each system call, then get the total $S = \{1, 2, ..., m\}$. The *source* system-call sequence *s* then will be represented as: $s = (x_1, x_2, x_3, ..., x_n)$, and the *target* as: $t = (y_1, y_2, y_3, ..., y_m)$, where $x_i, y_i \in S$.

We analyze in detail about how our sequence-to-sequence system-call prediction model working. As shown in Fig. 3, given a system-call sequence $(x_1, x_2, x_3, ..., x_n)$, the encoder completes a forward propagation operation to generate the hidden state and output at each moment in time. The formula is as follows:

$$h_t = sigm(w^{HX}x_t + w^{HH}h_{t-1}) \quad (2)$$

$$y_t = w^{YH}h_t \quad (3)$$

The hidden state $h_t$ of the last moment was token as the encoded information vector *c* of the input system-call sequence. We call it the context vector, which will be used as the initial hidden state of the decoder and will affect each moment of the output of the decoder.

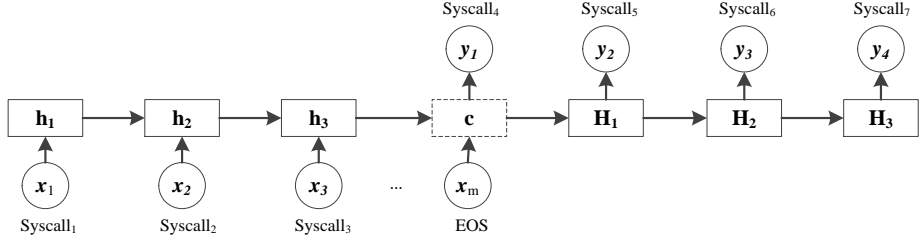

**Fig. 3.** the system-call sequence-to-sequence prediction model.

The decoder generates the *target* sequence *t* according to the context vector *c*, the decoder itself is essentially a standard recurrent neural network language model aforementioned, and the initial hidden state of the decoder is the context vector *c*. The process of generating each system-call requires the participation of hidden state $H_t$, also the system-call $y_t$ decoded at the previous moment. The condition probability of the decoder as follows:

$$p(y_1, \ldots, y_{T'} | x_1, \ldots, x_T) = \prod_{t=1}^{T'} p(y_t | c, y_1, \ldots, y_{t-1}) \qquad (4)$$

Although the model architecture Fig. 3 shown possesses certain capability to mapping sequence, there still some shortcomings. First, the input sentence information is encoded into a fixed vector in the encoding stage, which leads to the loss of certain information, the system-calls lying in the back position of input sequence may dilute the information of the previous ones. Second, the context vector *c* relied during decoding process is constant. That is to say, each system-call in the *source* sequence is viewed as the same without emphasis in the process of decoding.

In fact, we think that some specific system-calls will better reflect the program's intentions. Therefore, in order to emphasize the impact of those system-calls to the specific system-call decoding, we introduce the attention mechanism [23] into our system-call sequence prediction model.

The model with attention mechanism introduced has no change during the encoding stage. In the decoding stage, it will no longer rely on the context vector *c*, but instead take the information of the inputs which are most contributed when decoding each system-call. The model architecture with attention was shown in Fig. 4. Then the conditioned probability formula of decoder changed:

$$p(y_i | y_1, \ldots, y_{i-1}, X) = g(y_{i-1}, s_i, c_i) \qquad (5)$$

where $c_i$ stand for the current context information vector which calculated with attention algorithm learned during training and is calculated as the follows:

$$c_i = \sum_{j=1}^{T_x} a_{ij} h_j \qquad (6)$$

where $a_{ij}$ is normalized coefficient which weight each hidden state $h_j$ of time step $j$ contribute to the current context vector $c_i$ and $a_{ij}$ is calculated as:

$$a_{ij} = \frac{\exp(e_{ij})}{\sum_{k=1}^{T_x} \exp(e_{ik})} \qquad (7)$$

where $e_{ij}$ stands for the association relationship between the input element $x_i$ and the current output element $y_i$, which should be learned during the training phase.

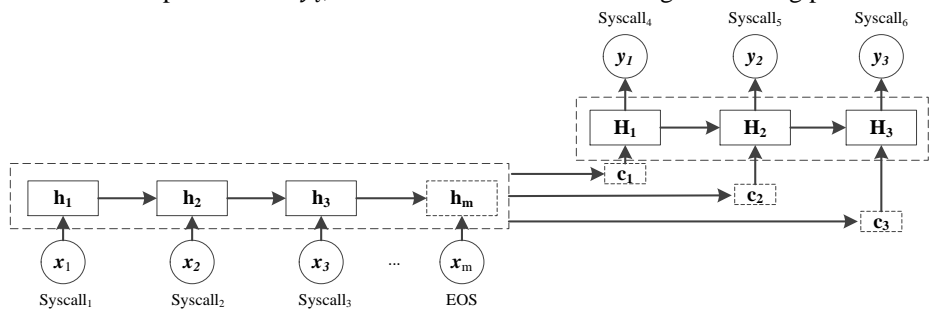

**Fig. 4.** the system-call sequence-to-sequence prediction model with attention.

By introducing attention mechanism, we enable the model to emphasize the influence of a particular system-call on the nature of a sequence, so it can better predict the actions that are most likely to occur in the next stage of the call sequence.

The performance of RNNs is relevant to the scale of training data set. In fact, because the abnormal system-call traces are less than the normal, the training input data constructed with ADFA-LD is slightly insufficient. For example, if we choose a sequence length of 25, we can only gain around 5000 pairs of input sequences. The approach we take to augment training data is to construct different training sequences by adjusting the length of system-call traces. We construct data set with sequence length of 10, 12, 15, 18, 20, 22, 25 and 30, respectively. In the end, we got more than 60,000 training data. On the other hand, the system-call traces during an intrusion period are confused with some other processes, the start and the termination boundary of a certain process are ambiguous. Therefore, it is necessary to take a longer sequence to extract the attack characteristics. By changing the length of the sequence, we can better capture the characteristics of the specific running process and achieve better predict results.

**Table 1.** training system-call traces generated with various length

| Train | | Test | |
|---|---|---|---|
| Traces length | Pairs | Traces length | Pairs |
| 10 | 8000 | 10 | 1600 |
| 12 | 7500 | 12 | 1500 |
| 15 | 7000 | 15 | 1400 |
| 18 | 6500 | 18 | 1210 |

| 20 | 6000 | 20 | 1200 |
|----|------|----|------|
| 22 | 5500 | 22 | 1100 |
| 25 | 5000 | 25 | 1000 |
| 30 | 4000 | 30 | 800  |

## 4  Experiment and Evaluation

We evaluate our model on our constructing evaluating data set. we apply the BELU [24] score, a benchmark widely used in machine translation, to evaluate our model performance. Also, we calculate the Euclidean distance of the encoded semantic vector between the model output and validation. In addition, the TD-IDF algorithm is employed in our work to evaluate the similarity between the predicted sequence and the validation. For comparison, we also train the standard recurrent neural network language model to generate a predictive sequence that compares with our predictive model.

In general, there are differences between the sequence predicted and the sequence of the real system-call. The system-call traces are affected by the processes running on the system, the hardware of the services, and the duration of the intrusion activities. However, it can be determined that the properties of system-call sequence predicted for a particular attack is consistence with the *target* sequence. To verify that our model predicts sequences with the same functionality of the *target*, we employ multiple anomaly detection classifiers to evaluate the predicted sequences, the results denote that our model well predicted the follow-up actions of those intrusion programs. What's more, we analyze the sequences which are composed of the predicted and the invoked to enhance the performance of anomaly intrusion detection, which means that our model has the ability to foresee the upcoming intrusions.

### 4.1  Dataset

The training data set is constructed on the ADFA-LD [25] provided by National Defense University of Australia, which consists of system-call traces gathered during normal and intrusion programs running. There are 746 files containing 6 types of attacks, 833 normal training files, and 4372 normal validation files.

**Table 2.** overall description of the ADFA-LD traces files

| Type | Traces files |
|------|--------------|
| Normal train | 833 |
| Normal validation | 4373 |
| Hydra_FTP | 162 |
| Hydra_SSH | 176 |
| Adduser | 91 |
| Java_Meterpreter | 124 |
| Meterpreter | 75 |
| Web_Shell | 118 |

Compared to the early UNM and DARPA data sets, ADFA is the latest, the attack pattern of experiment is more in line with the modern attack means, which indicates that the data is more consistent with the actual network environment. And the scale of the data set is large enough to be suitable for training neural network. The ADFA-LD data set is mainly to solve the problem which the early intrusion detection evaluation data set is not consistent with actual network environment. Creech *et al.* [25] introduces the structure, generation and experiment process of the ADFA-LD in detail, then their subsequent work [26] proposed a semantic approach to anomaly intrusion detection system using system-call patterns. And in [27] they proposed algorithm to build a host-based anomaly intrusion detection system, and preliminarily analyzed the data set.

### 4.2  Experiment Setup

In the training of neural networks, the selection of hyper-parameters often plays a critical role in the model convergence. The hyper parameters include:

1. the initialization of training model parameters,
2. learning rate,
3. batch size for min-batch optimization,
4. dropout applied to regularize model,
5. the number of neural nodes in the hidden layer of neural network.

There is no theoretical method to guide the determination of hyper parameters in the literature regarding deep learning. In general, there is a certain kind of experience to determine the appropriate hyper parameters for the model. For example, in the initialization of the parameters, the smaller random numbers distributed between 0 and 1 are generally chosen, and the learning rate can be decreased during the training iteration.

In our experiment, four models with various structures are trained as prediction models, including GRU unit of single hidden layer, two hidden layers and three hidden layers with learning rate of 0.001, and three hidden layers with learning rate of 0.1 respectively. The number of neural nodes in hidden layers of each model is 256, and the input dimension keeps the same with the hidden layer, which is due to the output of the GRU hidden layer is the input of the next layer. At the same time, the dropout is set to 0.5 to regularize the model. The maximum gradient clipped threshold is set to 5. The size of the training batch is 64. We adapt the early stop method to train models, when the current training loss no longer declines, the training is completed. The bucket strategy is employed to put the similar length sequence into the same size bucket, so as to reduce the paddings of training sequence and make the neural network training more efficient.

### 4.3 Result Analysis

First, we use the BELU score to evaluate our trained model performance. We use the *target* sequence in the test set as reference sequence. The BELU of the prediction sequence is calculated for different models of the hyper parameters.

**Table 3.** BELU scores of the four models with different structures

| Model | BELU score |
|---|---|
| GRU-1, lr=0.1 | 20.2 |
| GRU-2, lr=0.1 | 38.5 |
| GRU-3, lr=0.1 | 40.5 |
| GRU-3, lr=0.001 | 40.4 |

Table 3 denotes that when the learning rate keeping the same, the difference of BELU score between the number of hidden layers in the model is huge, which demonstrates that the increase of hidden layers can result in better performance. On the other hand, we can see that the difference of the initial learning rate in this experiment did not much affect the convergence of the model. It can be explained that we adopted the strategy that dynamically reduces the learning rate during training phase according to the training loss. According to the results, we can set a relatively large learning rate to accelerate the convergence speed of the model and reduce the training time.

In addition, we train various lengths of input sequences to enhance the robustness of the model. We classify the sequence according to the length and calculate the BELU score.

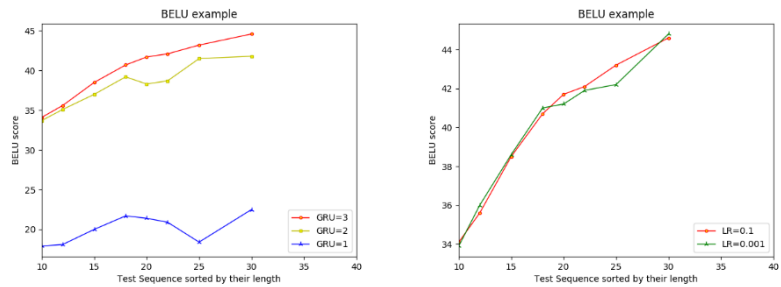

**Fig. 5.** the BELU score calculated with various lengths of predicted sequence. Left comparing the models with 1, 2, 3 hidden layers and learning rate keep in 0.001. Right comparing the models with the same 3 hidden layers while learning rate 0.1 and 0.001 respectively.

According to Fig. 5, we can see that when the sequence length is within 35, the value of BELU is positive proportion to the input length. It can be explained that our prediction model obtains more effective information by increasing the length of the input sequence, so that the results can be predicted more accurately. When the sequence length continues to grow, the BELU value will decline, which is due to the limitation of GRU memory, the system-call back position in the sequence dilute the information of the previous system calls causing the loss of information, and then the prediction errors increased. Therefore, when using the model, one need to choose appropriate length of the system-call sequence as input. In the other hand, we know that the influence of neural network depth on the model is huge.

In fact, the BELU indicators consider only the number of co-occurrences of N-tuples in the predicted sequence and the *target* sequence. The semantic information of the sequence is not considered. Generally speaking, there are key words existed in a sentence. Similarly, we deem that some specific sequences of system-call determine the characteristic of certain sequence. The TF-IDF (term frequency-inverse document frequency) model which widely used in information retrieval and data mining can well help us find the system calls that contribute to identify the sequence. This model extracts the key words of the statement by weighting the sequence specific words. We measure the similarity between the predicted sequence and the *target* sequence by this method. Fig. 6 (a) demonstrates that the predicted sequence well coincides with the *target* sequence in some syntactic level.

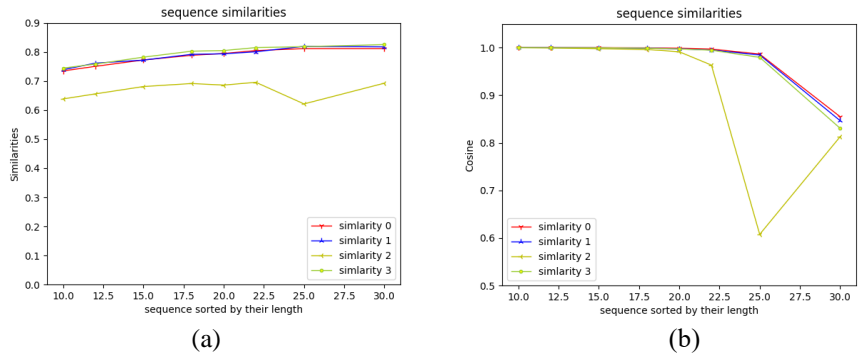

**Fig. 6.** The left represents the sequence similarity calculated using TF-IDF model. The right represents the semantic similarity calculated the cosine value of the two encoded vectors. curves represent different hyper-parameter models.

The disadvantage of both BELU and TF-IDF is that only the text similarity is considered, BELU considers the number of N tuples in the two sequence, and TF-IDF considers the similarity of the sentence key words. However, the similarity measure of sequence semantics can better evaluate the correlation of the two sequences. We take the final hidden state as the semantic encoding vector $v_1$ of prediction sequence, we do the same operation of the *target* sequence to get the corresponding semantic vector $v_2$. Then, we calculate the cosine value between the vectors. Based on the result of fig. 6, we demonstrate that the predicted system-call sequences of the model are very close to the *target* sequences both at the syntactic and the semantic level in certain range of prediction length.

One critical evaluation criterion of the predicted sequence is that it should keep the same functionality with the *target*, in other words, it should gain the similar performance tested under baseline anomaly detection classifiers. In order to verify our model, we evaluated the predicted sequence using classifiers including CNNs, RNNs, SVM and Random Forest. The ROC curve is applied to measure the detect performance of the predict sequence. Fig. 7 demonstrate that the predicted sequences show the remarkable performance towards those classifiers. Compare with the *target* sequence, the AUC score is with little decline of average 0.1, that means that our model well predicts the most likely system-call sequence with a high accuracy with regarding to sequence functionality.

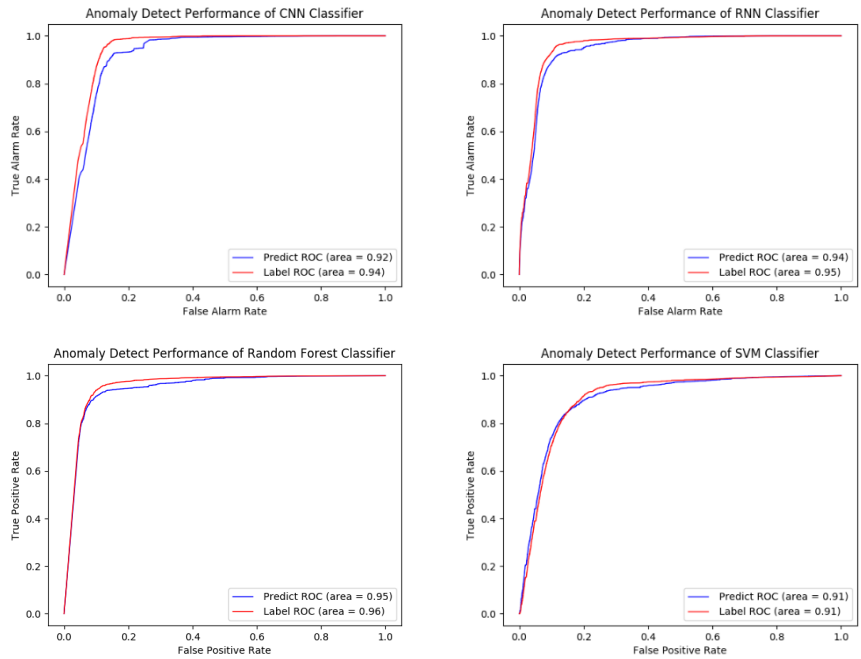

**Fig. 7.** ROC curves drawn with the four intrusion detection classifiers, blue line is evaluated on the *target* sequence, red line is evaluated on the predicted sequence

### 4.4 Enhancement of Anomaly Detection

There is another important benefit of predict model is that, with the predicted sequence, the anomaly detect system improve the intrusion detect ability with combining the original sequence.

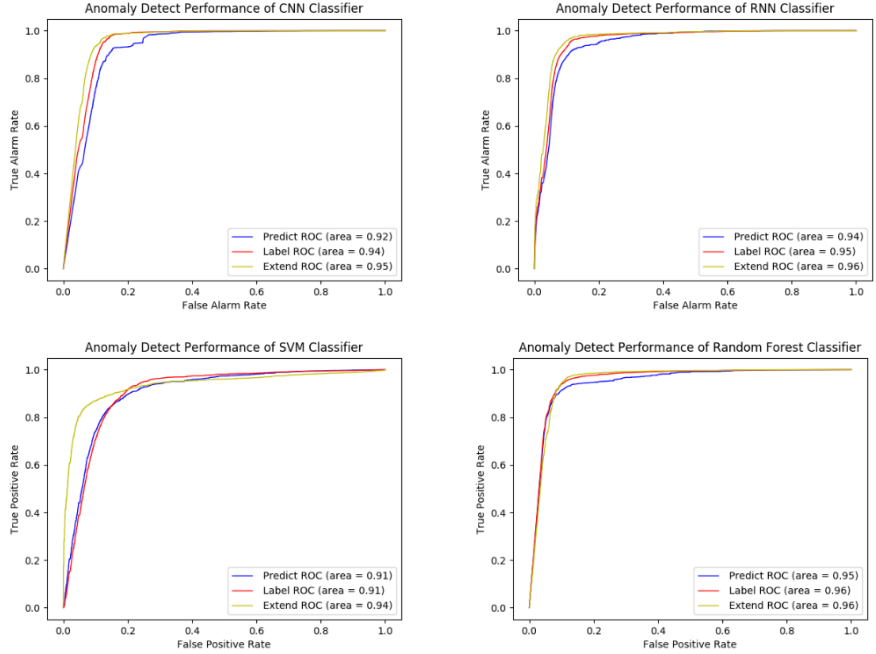

**Fig. 8.** ROC curves drawn with the four intrusion detection classifiers, blue line is evaluated on the *target* sequence, red line is evaluated on the predicted sequence, yellow line is evaluated on the extended sequence of the *target* and predicted.

We find from Fig. 8 that the extended sequence performs better than the original *target* sequence. We argue that this is because our prediction sequence captures the abnormal behavior characteristics of the existing sequence and predicts the sequence of subsequent behavior reasonably, thus helping the classification model to make better anomaly detection.

## 5 Conclusion

In order to solve the defect of the intrusion detection system's unable to predict intrusion behavior, the recurrent neural network sequence-to-sequence framework is introduced in our work. With the trained model, we can efficiently predict the reliable system-call sequence according to the invoked system-call sequence during the process running. The model is validated by evaluate the predict accuracy through various benchmark methods. On the other hand, by extending the system-calls sequence with the model predicted sequence we improve significantly the performance of anomaly detection algorithms. Future works will include enhancing the model robustness against adversarial samples and validating the portability performance on other data sets of our prediction model.

15# Reference

1. Dorothy E. An Intrusion-Detection Model[J]. IEEE Transactions on Software Engineering, 2006, SE-13(2):222-232.
2. Forrest S, Hofmeyr S, Somayaji A. The Evolution of System-Call Monitoring[C]// Computer Security Applications Conference, 2008. ACSAC 2008. IEEE, 2008:418-430.
3. Debar H, Becker M, Siboni D. A Neural Network Component for an Intrusion Detection System[C]// IEEE Symposium on Security and Privacy. IEEE Computer Society, 1992:240.
4. Schmidhuber J. Deep Learning in neural networks: An overview.[J]. Neural Networks the Official Journal of the International Neural Network Society, 2015, 61:85-117.
5. Xu Z, Yu X, Tari Z, et al. A multi-module anomaly detection scheme based on system call prediction[C]// Industrial Electronics and Applications. IEEE, 2013:1376-1381.
6. Mikolov T, Karafiát M, Burget L, et al. Recurrent neural network based language model[C]// INTERSPEECH 2010, Conference of the International Speech Communication Association, Makuhari, Chiba, Japan, September. DBLP, 2010:1045-1048.
7. Sutskever I, Vinyals O, Le Q V. Sequence to Sequence Learning with Neural Networks[J]. 2014, 4:3104-3112.
8. Shah B, H Trivedi B. Artificial Neural Network based Intrusion Detection System: A Survey[J]. International Journal of Computer Applications, 2012, 39(6):13-18.
9. Staudemeyer R C, Omlin C W. Evaluating performance of long short-term memory recurrent neural networks on intrusion detection data[C]// South African Institute for Computer Scientists and Information Technologists Conference. 2013:218-224.
10. Kim G, Yi H, Lee J, et al. LSTM-Based System-Call Language Modeling and Robust Ensemble Method for Designing Host-Based Intrusion Detection Systems[J]. 2016.
11. Feng L, Guan X, Guo S, et al. Predicting the intrusion intentions by observing system call sequences ☆[J]. Computers & Security, 2004, 23(3):241-252.
12. Zhang Z, Peng Z, Zhou Z. The Study of Intrusion Prediction Based on HsMM[C]// Asia-Pacific Services Computing Conference, 2008. APSCC '08. IEEE. IEEE, 2009:1358-1363.
13. Qiao Y, Xin X W, Bin Y, et al. Anomaly intrusion detection method based on HMM[J]. Electronics Letters, 2002, 38(13):663-664.
14. Zhitang Li, Lei J, Wang L, et al. A Data Mining Approach to Generating Network Attack Graph for Intrusion Prediction[C]// International Conference on Fuzzy Systems and Knowledge Discovery. IEEE, 2007:307-311.
15. Werbos P J. Backpropagation through time: what it does and how to do it[J]. Proceedings of the IEEE, 1990, 78(10): 1550-1560.
16. Bengio Y, Frasconi P, Simard P. The problem of learning long-term dependencies in recurrent networks[C]// IEEE International Conference on Neural Networks. IEEE, 1993:1183-1188 vol.3.
17. Bengio Y, Simard P, Frasconi P. Learning long-term dependencies with gradient descent is difficult[J]. IEEE Trans Neural Netw, 2002, 5(2):157-166.
18. Hochreiter S, Schmidhuber J. Long short-term memory.[J]. Neural Computation, 1997, 9(8):1735-1780.
19. Cho K, Van Merrienboer B, Gulcehre C, et al. Learning Phrase Representations using RNN Encoder-Decoder for Statistical Machine Translation[J]. Computer Science, 2014.
20. Chung J, Gulcehre C, Cho K H, et al. Empirical Evaluation of Gated Recurrent Neural Networks on Sequence Modeling[J]. Eprint Arxiv, 2014.
21. Bengio Y, Vincent P, Janvin C. A neural probabilistic language model[J]. Journal of Machine Learning Research, 2006, 3(6):1137-1155.